\documentclass[a4paper]{spie} 
\usepackage[]{graphicx}
 
\title{DUNE: The Dark Universe Explorer} 

\author{Alexandre R{\'e}fr{\'e}gier\supit{a}, Olivier Boulade\supit{a}, Yannick Mellier\supit{b}, Bruno Milliard\supit{c}, Reynald Pain\supit{d}, Joel Michaud\supit{e}, Fr{\'e}d{\'e}ric Safa\supit{f}, Adam Amara\supit{a}, Pierre Astier\supit{d}, Etienne Barrelet\supit{d}, Emmanuel Bertin\supit{b}, S\'{e}bastien Boulade\supit{f}, Christophe Cara\supit{a}, Arnaud Claret\supit{a}, Laurent Georges\supit{f}, Robert Grange\supit{c}, Julien Guy\supit{d}, Charles Koeck\supit{f}, Laurent Kroely\supit{d}, Christophe Magneville\supit{g}, Nathalie Palanque-Delabrouille\supit{g},  Nicolas Regnault\supit{d}, G\'{e}rard Smadja\supit{h}, Carlo Schimd\supit{g} and Zhihong Sun\supit{i}
\skiplinehalf
\normalsize
\supit{a} Service d'Astrophysique, DAPNIA, CEA Saclay, 91191 Gif sur Yvette, France; \\
\supit{b} Institut dÕAstrophysique de Paris, UMR 7095, Universit\'{e} Pierre \& Marie Curie, 98 bis Bvd Arago 75014 Paris, France\\
\supit{c} Laboratoire d'Astrophysique de Marseille, 13376 Marseille, France; \\
\supit{d} Laoboratoire de Physique Nucleaire et des Hautes Energies, IN2P3/CNRS et Universite Paris VI et VII,  4, Place Jussieu 75252 Paris Cedex 05, France\\
\supit{e} Centre National d'Etudes Spatiales, 31401 Toulouse, France; \\
\supit{f} EADS Astrium, 31402 Toulouse, France; \\
\supit{g} Service de Physique des Particules, DAPNIA, CEA Saclay, 91191 Gif sur Yvette, France; \\
\supit{h} Institut de Physique Nucleaire de Lyon, CNRS/IN2P3,Universite Cl. Bernard Lyon 1, 4 rue Enrico Fermi, 69622 Villeurbanne Cedex, France\\
\supit{i} Service d'Ing\'{e}nierie des Syst\`{e}mes, DAPNIA, CEA Saclay, 91191 Gif sur Yvette, France
}
\authorinfo{For further information contact: Alexandre R{\'e}fr{\'e}gier, email: refregier@cea.fr}

\begin{document}

\maketitle 

%%%%%%%%%%%%%%%%%%%%%%%%%%%%%%%%%%%%%%%%%%%%%%%%%%%%%%%%%%%%% 
\begin{abstract}
Understanding the nature of Dark Matter and Dark Energy is one of the most 
pressing issues in cosmology and fundamental physics.
The purpose of the DUNE (Dark UNiverse Explorer) mission is
to study these two cosmological components with high precision,
using a space-based weak lensing survey
as its primary science driver. Weak lensing provides a measure of the distribution of dark matter in the universe and of the impact of dark energy on the growth of structures. DUNE will also include a complementary
supernovae survey to measure the expansion history of the universe, thus
giving independent additional constraints on dark energy.
The baseline concept consists of a 1.2m telescope with a 0.5 square
degree optical CCD camera. It is designed to be fast with reduced risks and costs,
and to take advantage of the synergy between ground-based and space observations.
Stringent requirements for weak lensing systematics  were shown to be achievable with the baseline concept. This will allow DUNE to place strong constraints on cosmological parameters, including the equation of state parameter of the dark energy and its evolution from redshift 0 to 1. DUNE is the subject of an ongoing study led by the 
French Space Agency (CNES), and is being proposed for ESA's Cosmic Vision programme. 
\end{abstract}

%>>>> Include a list of keywords after the abstract 


%%%%%%%%%%%%%%%%%%%%%%%%%%%%%%%%%%%%%%%%%%%%%%%%%%%%%%%%%%%%%
\section{INTRODUCTION}
\label{sect:intro}  % \label{} allows reference to this section

The recent measurements of Cosmic Microwave Background (CMB)
anisotropies, Supernovae Ia, and large-scale structure concur to
confirm the $\Lambda$CDM cosmological model (see ref.~\citenum{spe06} and refs therein). Paradoxically, this
`concordance' model relies on three ingredients whose origin and
nature are unknown: dark matter (CDM), dark energy ($\Lambda$) and the
fundamental field(s) driving inflation. The understanding of these
ingredients is likely to revolutionalise fundamental physics. Recent ground-based observations (see refs. \citenum{eis05,hoe06,sem06,ast06} and refs. therein) have started to constrain the properties of dark energy and dark matter. Because the effect of dark energy on these two probes are very small, ground based surveys will however eventually be limited by systematics which can be circumvented by wide-field space observations.  

The DUNE (Dark UNiverse Explorer) mission is a wide-field space mission concept aimed at the high-precision study of the properties dark matter and dark energy. 
For this purpose, DUNE will use the weak lensing technique to measure the distribution of dark matter in the universe and the impact of dark energy on the growth of structure. DUNE will also include a complementary Supernovae survey to measure the expansion history of the universe, thus giving further constraints on dark energy, independent of those from weak lensing. The DUNE concept builds upon instrumental experience from the MegaCam\cite{ccds} camera installed at the Canada-France-Hawaii telescope and upon the weak lensing\cite{hoe06,sem06} and supernovae\cite{ast06} measurements recently performed with this instrument. DUNE is the object of an ongoing study phase led by the French Space Agency (CNES), and is being proposed for ESA's Cosmic Vision programme. 

In these proceedings, we describe the mission concept developed in the course of the pre-study (phase 0) study led by CNES in 2005. After presenting the science goals and the science requirements (\S\ref{science}), we describe the baseline configuration for the instrument (\S\ref{instrument}), and for the spacecraft and mission (\S\ref{spacecraft}). Perspectives and a description of ongoing studies are described in the conclusion (\S\ref{conclusion}).

%%%%%%%%%%%%%%%%%%%%%%%%%%%%%%%%%%%%%%%%%%%%%%%%%%%%%%%%%%%%%
\section{Science Case}
\label{science}

\begin{figure}
\centering
\includegraphics[width=80mm]{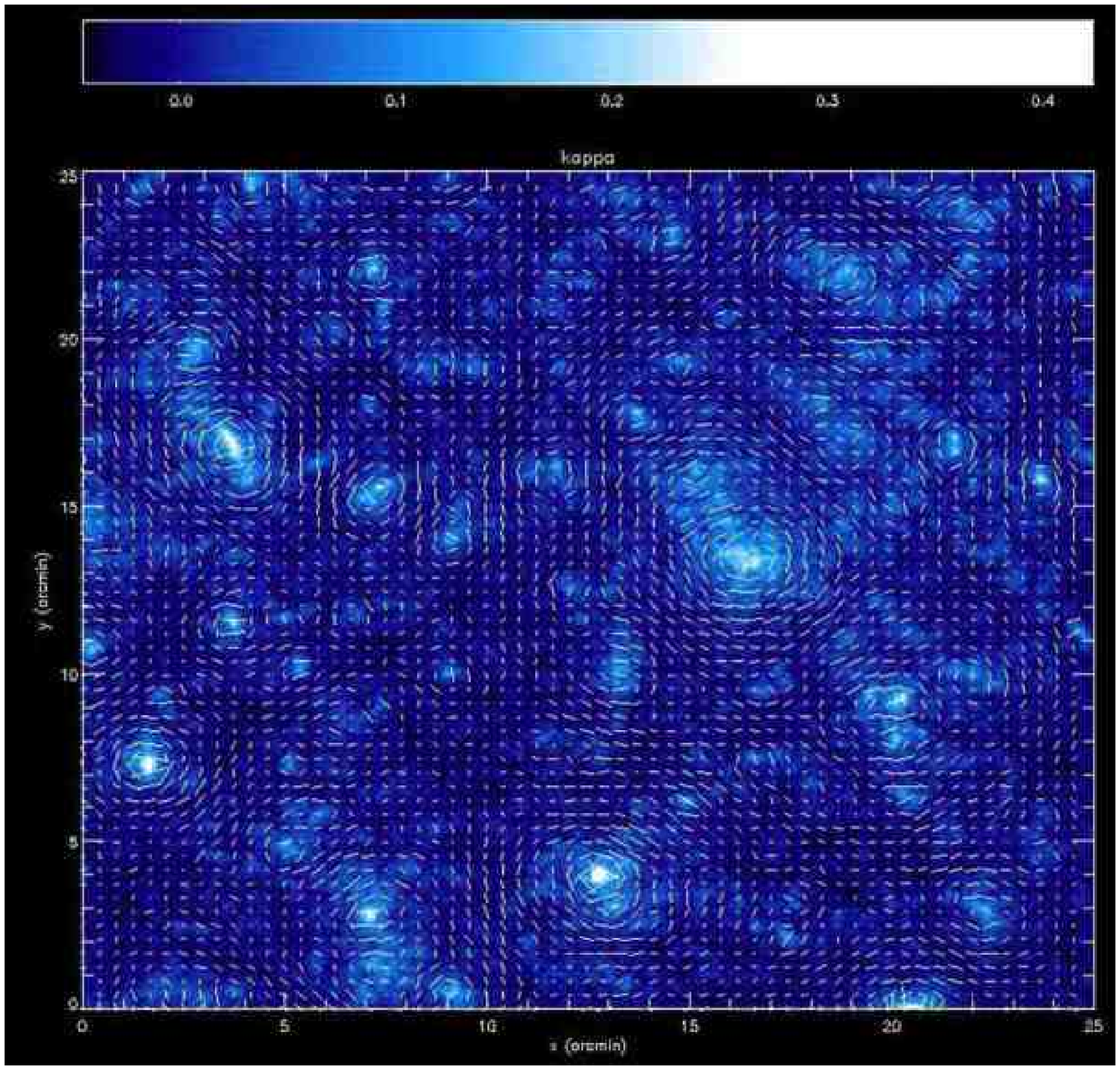}
\includegraphics[width=90mm]{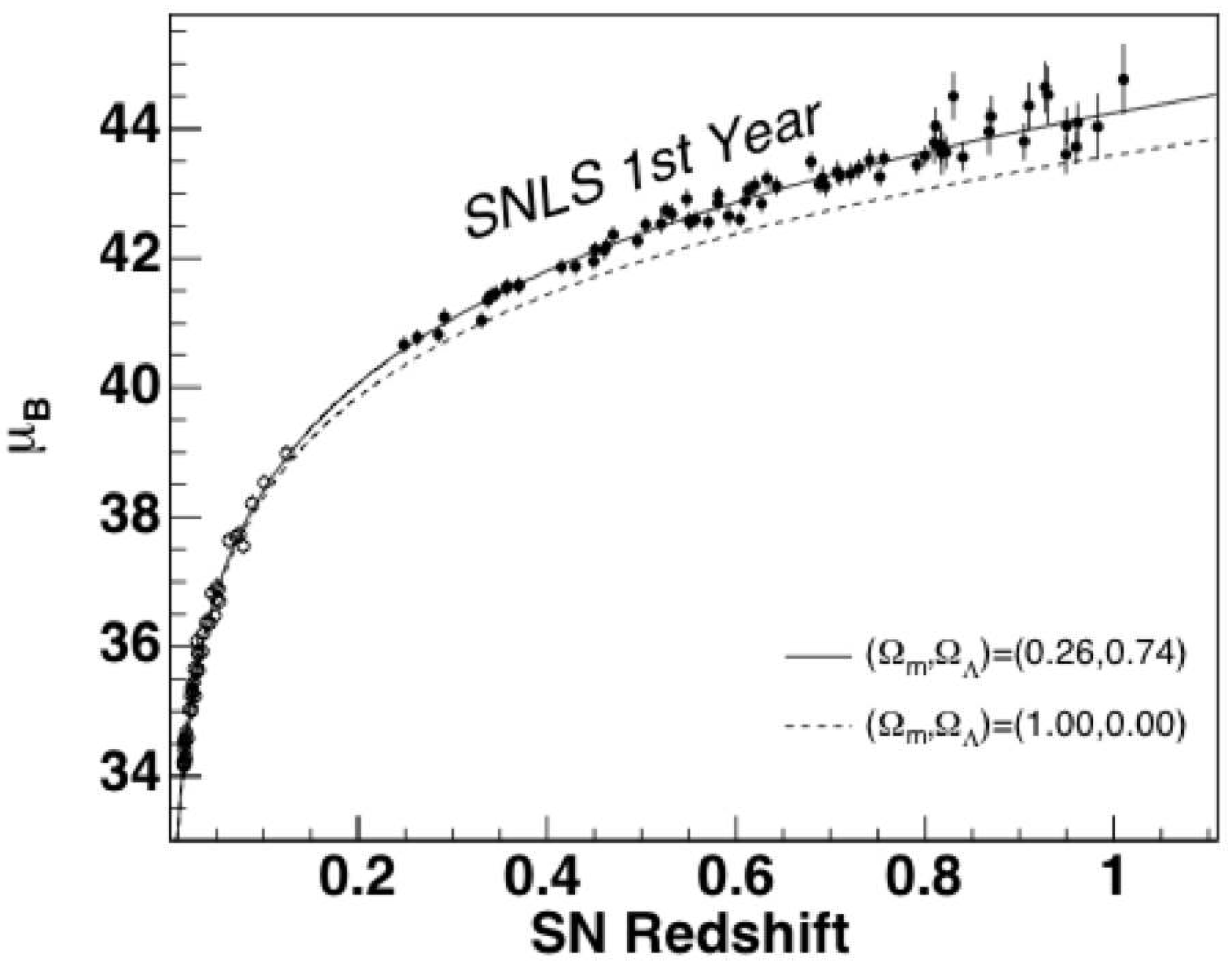}
\caption{\small The two cosmological probes used by DUNE. Left: Weak lensing field
for a region of the sky in a CDM model simulated by Jain et al.\cite{jai00}. The light (dark) regions corresponds to projected over(under)-densities of the dark matter. The line segments represent the shear to which the images of background galaxies are subject.
 Right: Hubble diagram for the first year results of the SNLS survey\cite{ast06}
 giving the relation between the apparent magnitude and the redshift
 of Type Ia supernovae. The predictions for two  cosmological models, with and
 without a cosmological constant $\Lambda$, are shown and illustrate the sensitivity of this technique to dark energy.
\label{fig:probes}}
\end{figure}

The purpose of DUNE is to perform a high-precision study of dark matter and dark energy using weak lensing and type Ia supernovae, two complementary cosmological probes. The weak lensing technique relies on the measurement of the small distortions induced by intervening large-scale structures on the images of distant galaxies (see refs.~\citenum{van03, ref03} for reviews).  These distortions can  be used to directly map the distribution of Dark Matter, which dominates the mass  in the universe, and thus to study dark energy, which affects the growth of structure (see Figure~\ref{fig:probes}a). Weak lensing measurements can also be used to measure the primordial   power spectrum of density  perturbations and thus, in combination with CMB measurements,  constrain models of inflation.  

Supernovae Ia form a homogeneous class of very luminous objects that have been shown to provide excellent distance indicators (see ref.~\citenum{lei01} for a review) to
probe the expansion rate of the universe. The relation between the apparent magnitude
and the redshift of the supernovae can thus be used to constrain the geometry of the
Universe and the amount and properties of dark energy (see Figure~\ref{fig:probes}b). 

The following describes the science goals and requirements of the DUNE mission for each of these probes, along with simulations studies showing how these goals can be achieved with the DUNE baseline configuration.

\subsection{Science Goals}

The primary scientific goals of the DUNE mission are: (1) to measure the equation of state parameter $w$ of the dark energy and its evolution $w_a$ with a precision better than 5\%  and 20\%, respectively, in the redshift range $z=0$ to $1$. This will allow us to distinguish a cosmological constant model from alternative models, like quintessence with a dynamical scalar field. 
(2) To measure the mass power spectrum and higher order statistics along with their evolution from $z=0$ to $1$ from linear to non-linear
scales. This will allow us to test the dark
matter driven gravitational instability paradigm of cosmic structure
formation.
(3) to reconstruct the primordial matter power spectrum and constrain theories
of inflation by their effect on the shape of the present-day dark matter
power spectrum. The first science goal is best achieved using a combination
of weak lensing and supernovae measurements, whereas the final two rely on
measurements of large scale structure with weak lensing.

The space-based  wide field surveys required to achieve the primary goals of DUNE will also provide a wealth of secondary science returns. In particular, DUNE  will provide a mass-selected cluster catalogue out to $z \sim 1$,
a detailed description of the relation between light from galaxies and dark matter as function of redshift and scale, and a unique wide, high spatial resolution survey of galaxies.  DUNE will also yield a determination of the star formation rate as a function of redshift  through the measurement of core collapse supernovae rates  up to $z\sim 1$.  The combination of the DUNE weak lensing and supernovae surveys will also provide fundamental tests of the distinction between dark energy and the modification of gravity on large scales, the gravitational instability paradigm for structure formation, and for the presence of dark energy clustering.

\subsection{Science Requirements}
\label{requirements}
The instrumental and observing requirements needed to achieve the above scientific goals are summarised in table~\ref{tab:science}. To reach the necessary statistical precision, the weak lensing survey will cover 20,000 deg$^{2}$ at high galactic lattitude ($|b|>30^{\circ}$), thus avoiding the galactic plane. It will provide a sample of at least 35 galaxies per amin$^{2}$ usable for weak  lensing (i.e. with a SExtractor SNR$>7$ and a FWHM$>1.2\times$FWHM of the PSF) with a mean redshift $\langle z \rangle \simeq 1$ and an rms shear uncertainty per galaxy of $\sigma_{\gamma}=\langle |\gamma^2| \rangle=0.35$.  To significantly improve upon ground based observations, the PSF FWHM will be smaller than 0.23". A sampling of at least 2 pixels per PSF FWHM is required for proper PSF modeling and deconvolution. Photometric redshifts are needed to achieve three distinct redshift bins over the survey area and will be provided by multi-band photometry from a complementary ground based survey.

\begin{table}[htbp]
\caption{\label{tab:science}Summary of the science requirements} 
\begin{center}       
\begin{tabular}{|l|l|l|}
\hline
\hline
 & Weak lensing & Supernovae \\
\hline
sky coverage & 20,000 deg$^{2}$ ($|b|> 30^{\circ}$) 
 & $2\times60$ deg$^{2}$  yielding $\sim10,000$ SNe \\
 & ~~in blocks $>20\times20$ deg$^2$ & ~~with $0.1<z<1.1$ \\
image quality & PSF FWHM = 0.23 $''$  & idem \\
image sampling & 2 pixels per PSF FWHM & idem \\
bandpasses & 1 broad red band ($\sim$R+I+Z) & 6 bands$^a$ ($\sim$UBVRIZ)\\
time sampling & none & 6 bands  every 4 days for $2\times9$ months\\
sensitivity & $>$35 usable$^b$ galaxies/amin$^{2}$  with & SNR$>5$ for SNe at $z\simeq 1$\\
 & ~~$\langle z \rangle \simeq 1$ and $\sigma_\gamma=0.35$ per galaxy & \\
systematics & error in shear $< 0.1$\% (PSF $e<6$\%
& mean SNe luminosity error $<2$\% \\ 
& ~~and $\delta e<0.1$\% after calib.)& ~~in $\Delta z$=0.1 bin\\
ground-based complement &  multi-band photometry for photo-$z$'s & 
spectroscopy of host galaxies\\
\hline
\end{tabular}
\end{center}
{\small
$^a$ redshifted U- and B-bands\\
$^b$ galaxies usable for weak lensing, i.e. with SExtractor SNR $>$ 7 and FWHM $>$ 1.2$\times$FWHM[PSF] }
\end{table}

To reach the necessary precision on dark energy parameters, the systematic errors on the shear measurement after deconvolution of the PSF is required to be smaller than about 0.1\%. This corresponds roughly to an upper limit on the PSF ellipticity of $|e|<6\%$ accross the field of view with an uncertainty after calibration of $|\delta e|<0.1\%$. The survey region will be scanned in compact regions of at least $20^{\circ}$ on a side to allow measurement of the weak lensing signal on scales as large as $10^{\circ}$, even if the mission is prematurely interrupted, and to minimise systematics on smaller scales due to long term instrumental variations. Within a compact region, consecutive strips will be contiguous and will overlap by about 10\% for cross-calibrations of the photometry and of the PSF. Low cosmic ray levels, reduced stray light, linear and stable CCDs and achromatic optics must also be ensured to preserve good image quality. 

The supernovae survey will provide several hundred Type Ia supernovae per z=0.1 bin between$z\sim0.1$ and $z\sim1.1$. This requires surveying a total area of
$120$ deg$^{2}$  in 2 distinct regions ($2\times 60$ deg$^{2}$) in 6 bands for 9 months each. This will provide a total of about 10,000 Type Ia supernovae light curves. The restframe $U-$ and $B-$ band peak luminosity of the supernovae will be measured with an average 2\% statistical uncertainties. This requires photometric measurement in
redshifted $U$- and $B$-band (approximately UBVRIZ) of supernovae light curves every
4 days, from 2-3 weeks before maximum light to 3-4 weeks past maximum
light. The identification of the supernovae will be derived from the properties of their multi-color light curveys. Their redshift will be obtained from differed ground-based medium-resolution spectroscopy of the host galaxies.  

For the SNe survey, systematics such as malmquist bias, extinction by host galaxy, evolution of the supernovae luminosity and gravitational lensing must be controled to an
average level of $\sim2$\% per $z=0.1$ bin. This requires precise
photometry of the supernovae lightcurves in at least 3 bands from 2
weeks rest frame before maximum to about 6 weeks rest frame past
maximum.

\subsection{Simulations}
In order to estimate the sensitivity of the DUNE baseline for weak
lensing, we developped an image simulation pipeline based on 
{\tt skymaker}\footnote{available at {\tt http://terapix.iap.fr/clt/oldSite/soft/skymaker}} and 
adapted to DUNE. The simulation
include observational effects such as photon and readout noise, background light, and PSF convolution. The galaxy population parameters for the simulations have been calibrated against catalogues from the GOODS survey\cite{gia03}  observed with the ACS instrument on HST.  The simulations use the baseline instrument configuration listed in Table~\ref{tab:baseline} .

\begin{figure}
\centering
\includegraphics[width=80mm]{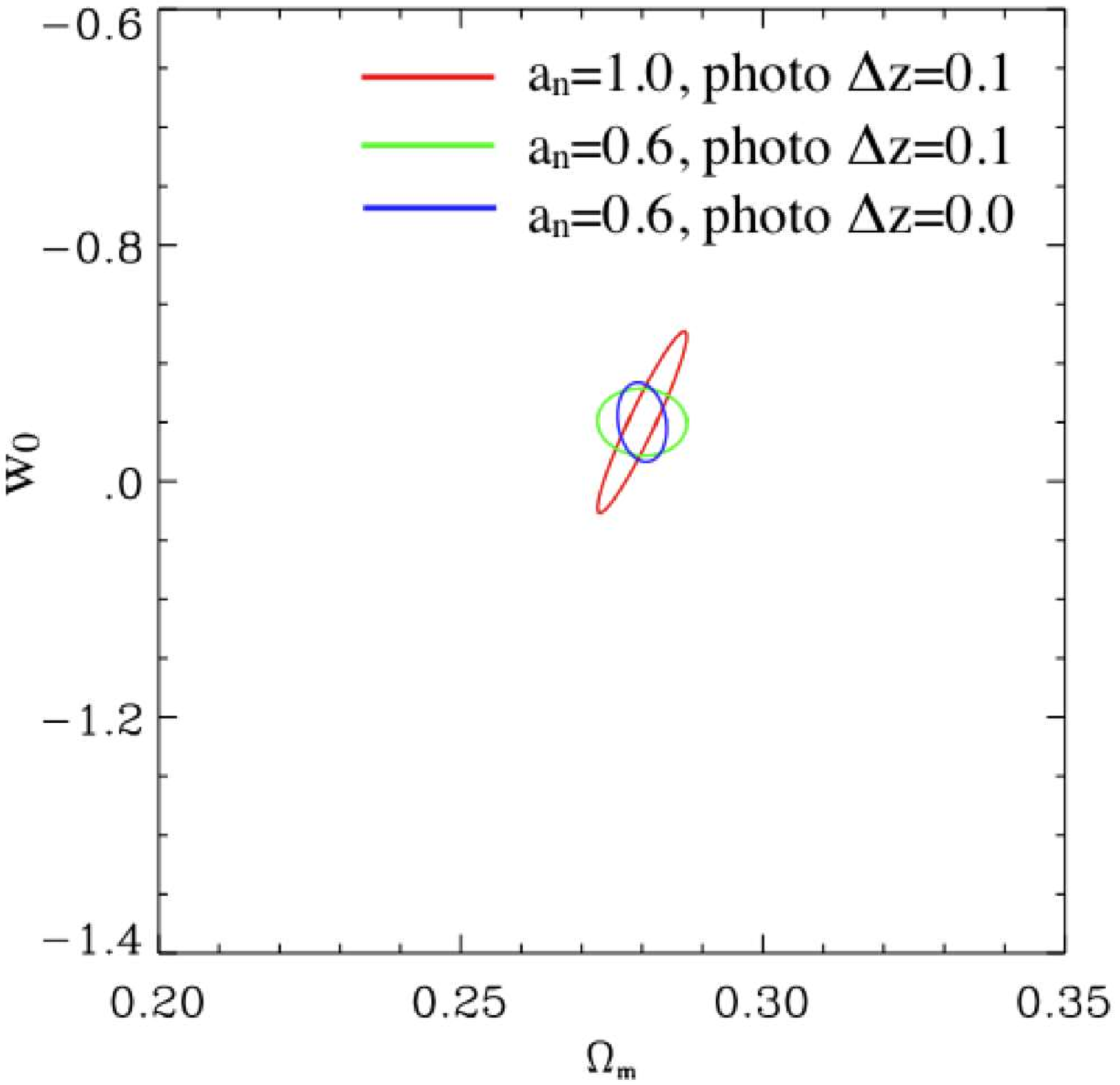}
\includegraphics[width=75mm]{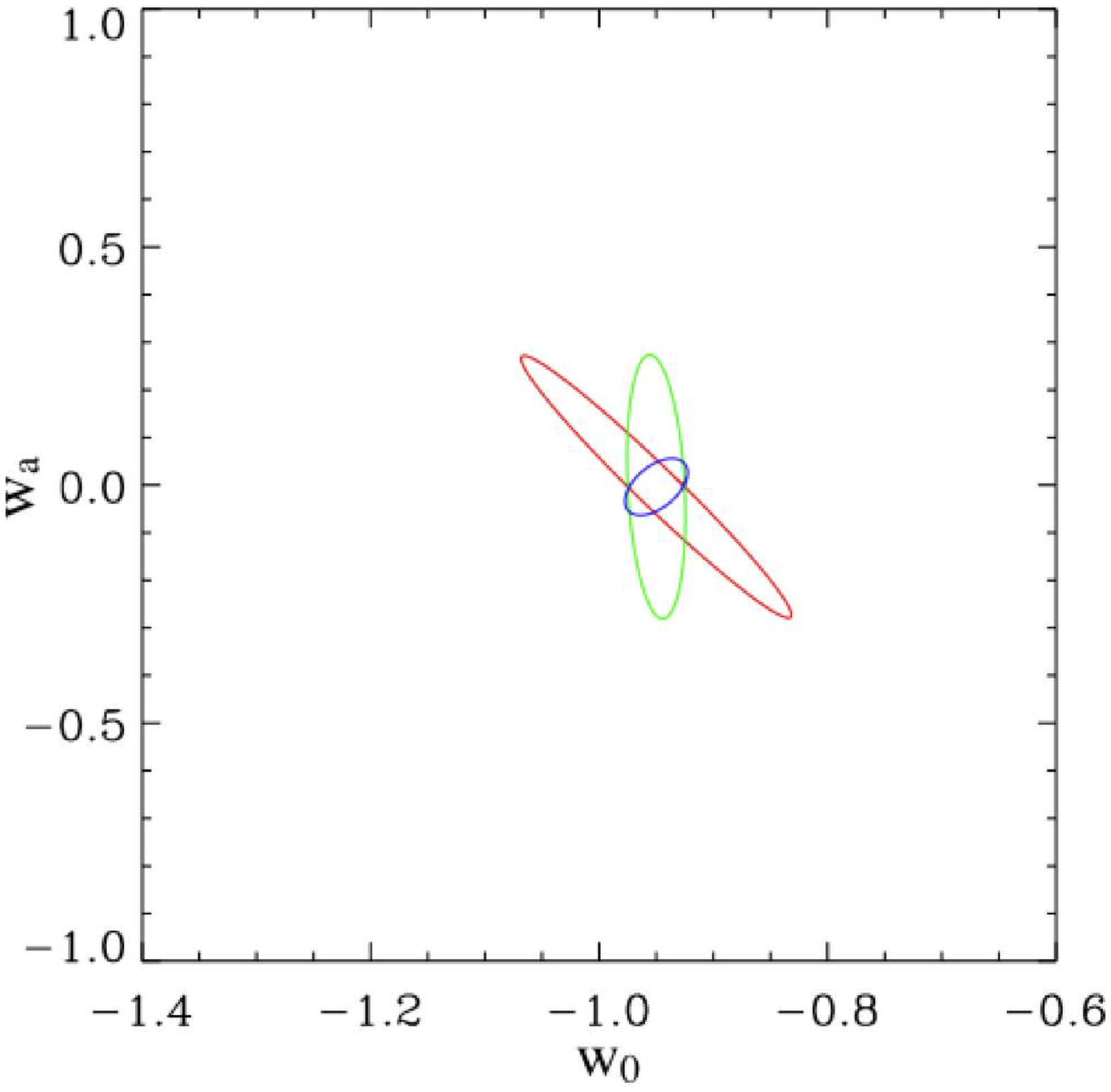}
\caption{Constraints on the cosmological parameters that can be achieved
by the DUNE weak lensing surveys as specified in the requirements.  The red, green and blue contours corresponds to different assumptions for the photometric redshift errors $\Delta z$ and for the pivot point $a_n$.
All contours correspond to 68\% confidence level from power spectrum tomography with
3 redshift bins, and are marginalised over 
the other five parameters in the set $[w_0,w_a,\Omega_m, \Omega_b, \sigma_8, n_s, h]$.
\label{fig:contours_wl}}
\end{figure}

\begin{figure}
\centering
\includegraphics[width=70mm,height=70mm]{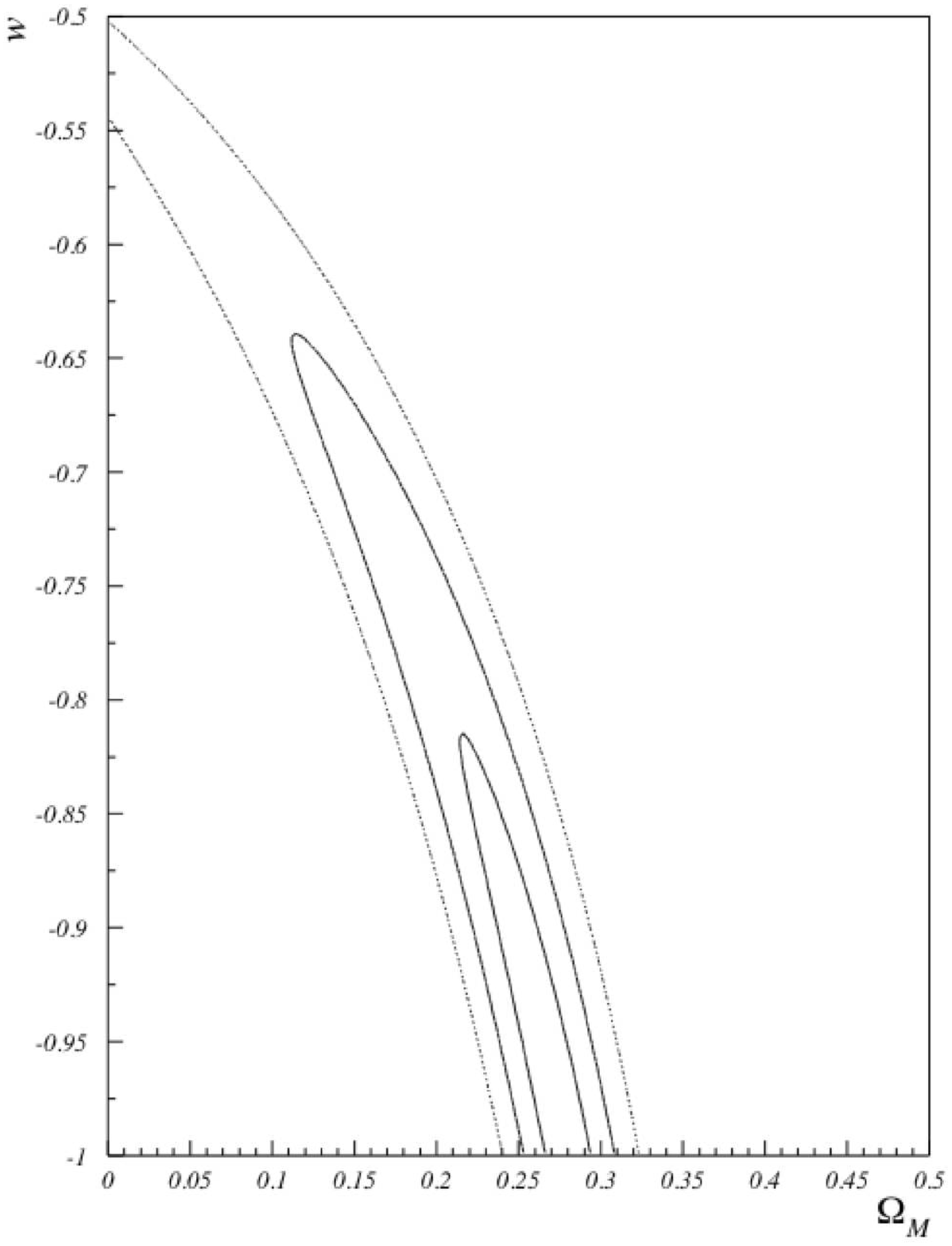}
\includegraphics[width=70mm,height=70mm]{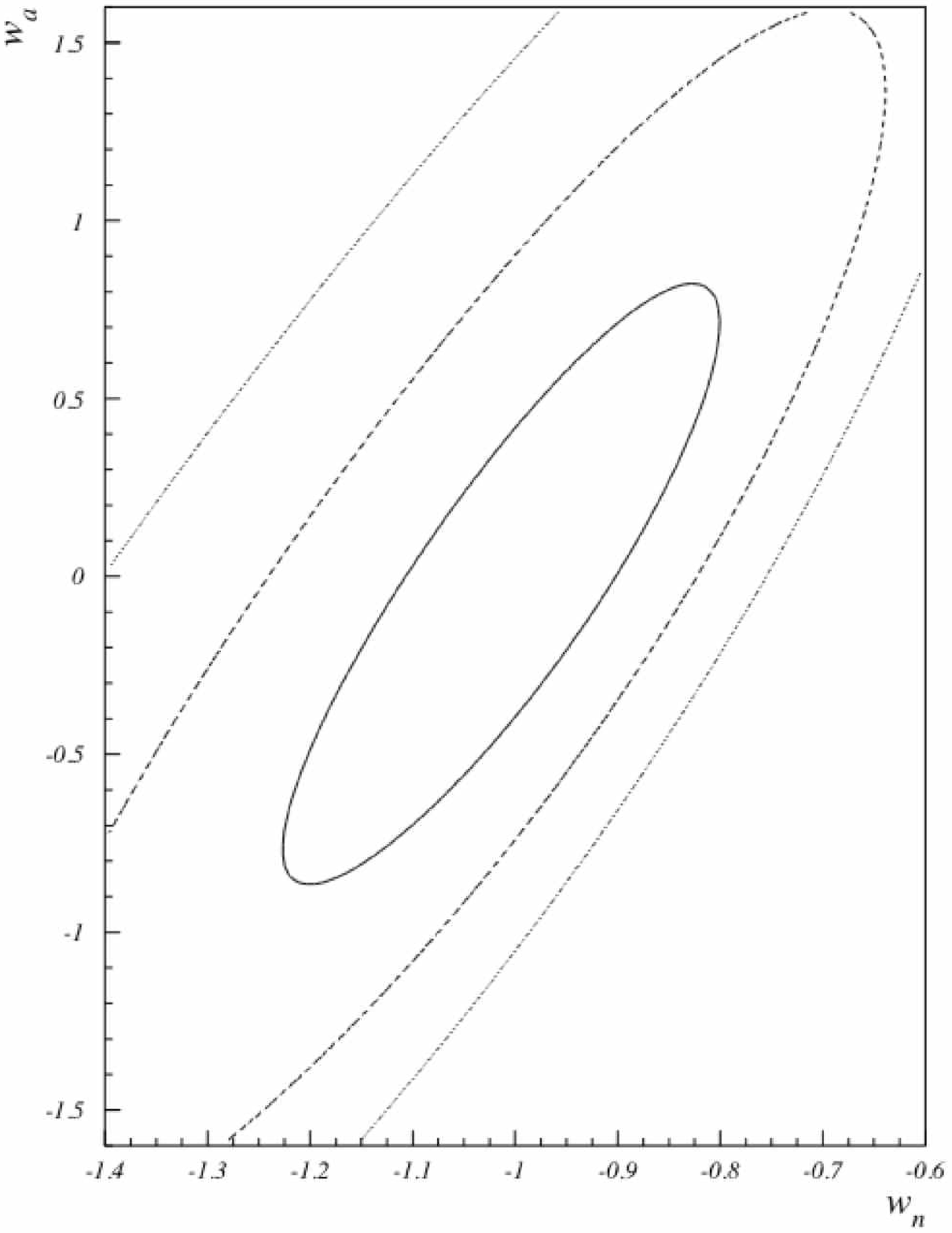}
\caption{Constraints on the cosmological parameters which can be derived
by the DUNE supernovae surveys as specified in the requirements. The  68\%, 95\% and 99\% confidence contours are displayed and include statistical and systematic errors (see text for details). Left: constraints in the ($\Omega_m$, $w$) plane for a flat universe
with no prior on $\Omega_m$, but assuming a constant $w$ ($w_a \equiv 0$).
Right: constraints in the $w_a, w_n$ 
plane for a flat universe with a prior of $\Omega_m=0.28\pm0.03$.
\label{fig:contours_sn}}
\end{figure}

Figures~\ref{fig:contours_wl} shows the accuracy achieved by the DUNE weak lensing
as estimated from the simulations and a fisher matrix calculation. 
The confidence contours  for the matter density parameter $\Omega_m$ and the dark energy parameters $w_n$ and $w_a$ are displayed using the weak lensing
power spectrum tomography with 3 redshift bins. We have used the
parametrisation\cite{che01,lin03} of the evolution of the dark energy equation of state parameter $w(a)=w_n+(a_n-a)w_a$ where $a_n$ is an adjustable pivot point
and $a=(1+z)^{-1}$ is the cosmic expansion parameter.
Figure~\ref{fig:contours_sn} shows confidence contours that can be achieved with the SNe survey. The latter contours include statistical and systematic uncertainties conservatively modelled as  $\delta m = 0.01(1+z)$, in magnitude per redshift bin of width 0.1. 

The image simulations were also used to study the trade off between a wide and deep survey
strategy for a fixed total survey time. For the purpose of measuring the dark
energy equation of state parameters ($w_n, w_a$) with weak lensing power spectrum
tomography, we found that a wide survey is more advantageous than a deep survey for a fixed amount of survey time. This gain justifies our choice for the survey strategy. The simulations have also shown the advantage of a wide
filter (covering the total range of R+I+z filters) to maximise
the galaxy surface density necessary for appropriate shear
measurements.  The impact of the finite size of
the PSF due to the optics has also been studied. In particular,
requiring a minimum SExtractor SNR of 7 for galaxies to be used
in shear measurements, a total (optics + diffraction off the primary)
PSF FWHM of 0.5$''$ instead of 0.23$''$ induces a loss of $\sim 50$\% on
the galaxy counts, roughly independently of the exposure time and filter. 
This implies a clear improvement of a space-based study over a ground-based experiment of the same depth, which have PSF FWHM of at least $~$0.5'' because of atmospheric seeing.

\section{Instrument Baseline} 
\label{instrument}
To meet the scientific requirements described in the previous section, we have designed a mission concept for a space-based wide field imager. The resulting baseline instrument and mission parameters  are summarized in table \ref{tab:baseline}. Some of these parameters (mission duration, size of the instrument field of view) are also derived from cost and feasability considerations. In particular, for the standard step-and-stare observing mode,
the time for the stabilization of the spacecraft at each pointing represents a sizeable fraction of the exposure time, making the observations inefficient. We have therefore chosen a Time-Delayed Integration (TDI) observing mode (or 'drift scan') in which the spacecraft scans long strips of the sky, while the CCD detectors shift the signal charges at the same rate, and which results in much higher observing efficiencies. In addition, the TDI observing mode avoids the need of a shutter, thus reducing the risks associated with this moving part.

\begin{table}[htbp]
\caption{\label{tab:baseline}Summary of the baseline instrument and mission parameters.} 
\begin{center}       
\begin{tabular}{|l|l|l|}
\hline
\hline
 & Weak lensing & Supernovae \\
\hline
filters & 1 band R+I+Z (783 $\pm$ 217 nm) & 6 bands (from 365 to 908 nm) \\
exposure time & 1500 seconds & 300 seconds \\
sky coverage & 20,000 deg$^2$ with $|b|> 30^{\circ}$ & 2 fields 60 deg$^{2}$ each\\
mission duration & 3 years (total with 60\% efficiency) & 2$\times$9 months (total with 60\% efficiency) \\
\hline
telescope size & \multicolumn{2}{c|}{1.2 m} \\
field of view & \multicolumn{2}{c|}{0.5 deg$^{2}$} \\
pixel scale & \multicolumn{2}{c|}{0.115 $''$} \\
observing mode & \multicolumn{2}{c|}{time delayed integration} \\
\hline
\hline
\end{tabular}
\end{center}
\end{table} 

\subsection{Optical design} 
\label{optics}

The baseline optical concept that fulfills the requirements for DUNE (called NODI for NO DIstortion), is based on a 3-mirror  Korsch\cite{kor77} type of telescope with an effective diameter of 1.2m.  We summarise here the main features of the optical concept (see ref.~\citenum{gra06} for a detailed description of the concept with slightly different parameters).
The combination of the primary and secondary mirrors is similar to a Cassegrain configuration and forms a real image just behind the primary. In contrast with the classical Korsch design where this intermediate image is re-imaged by a tertiary with a magnification close to unity, we explored solutions where a large magnification of the tertiary is used to reach the very low distortion required by the TDI scanning mode. For easy access to the focal plane, a folding flat mirror can be placed either near the exit pupil or between the intermediate image and the tertiary mirror. In the latter case, the folding mirror must be pierced to let the beam through. In these two configurations, the accessible focal plane is an annulus with an external diameter of about 640 mm.

This Korsch configuration has a number of advantages, the most important of which is 
that it provides an achromatic image limited by diffraction over a planar field of more than 1 deg$^2$ with 3 deformed mirrors (simple conical sections).
Another advantage is the excellent protection against stray light thanks to the positioning of the optical surfaces. An additional folding mirror also makes it possible to position the camera at 90 degrees from the optical axis close to the external envelope of the satellite, thus facilitating the cooling of the detectors.

\begin{figure}[htbp]
\begin{center}
\begin{tabular}{c}
\includegraphics[height=65mm]{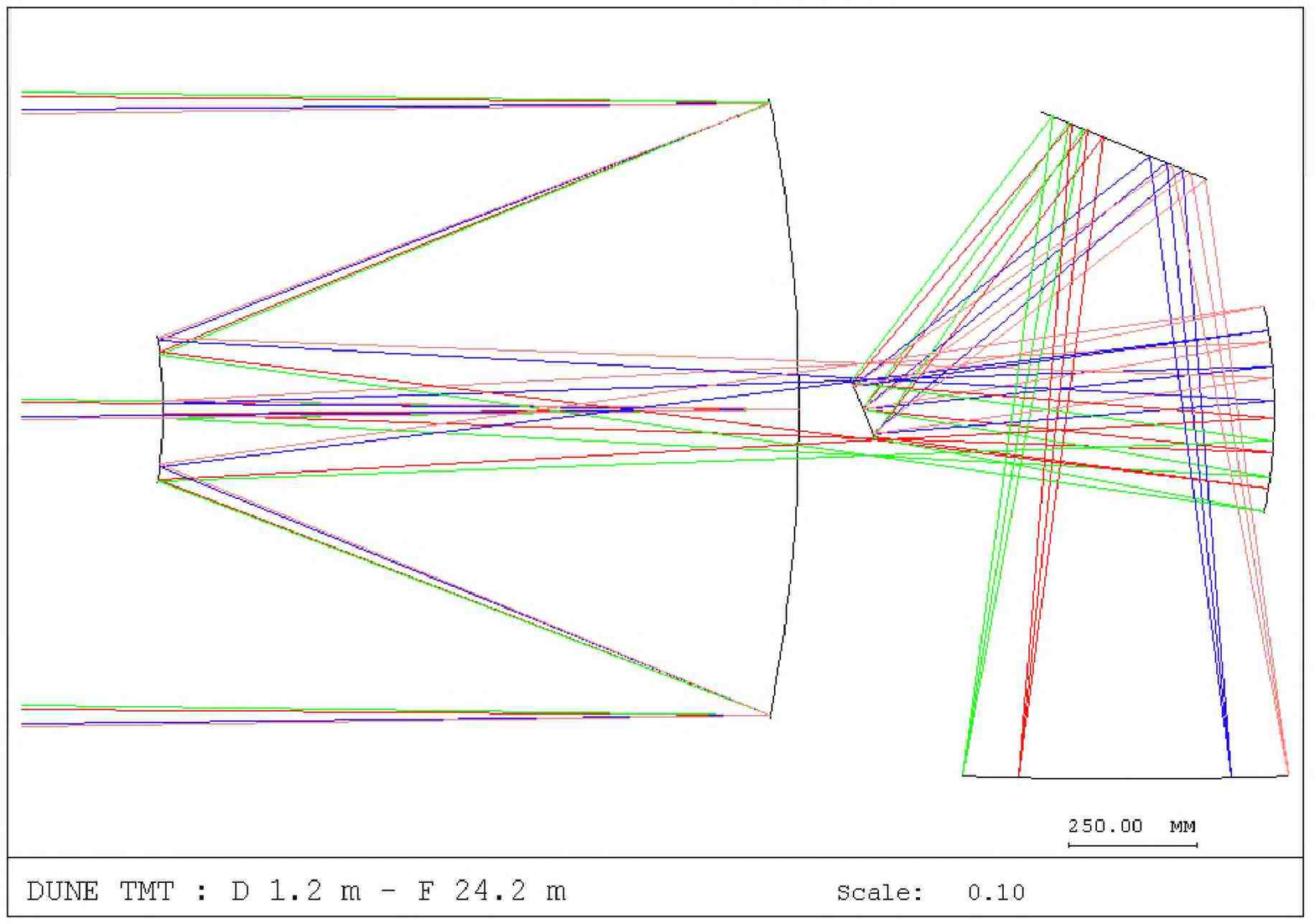}
\includegraphics[height=70mm]{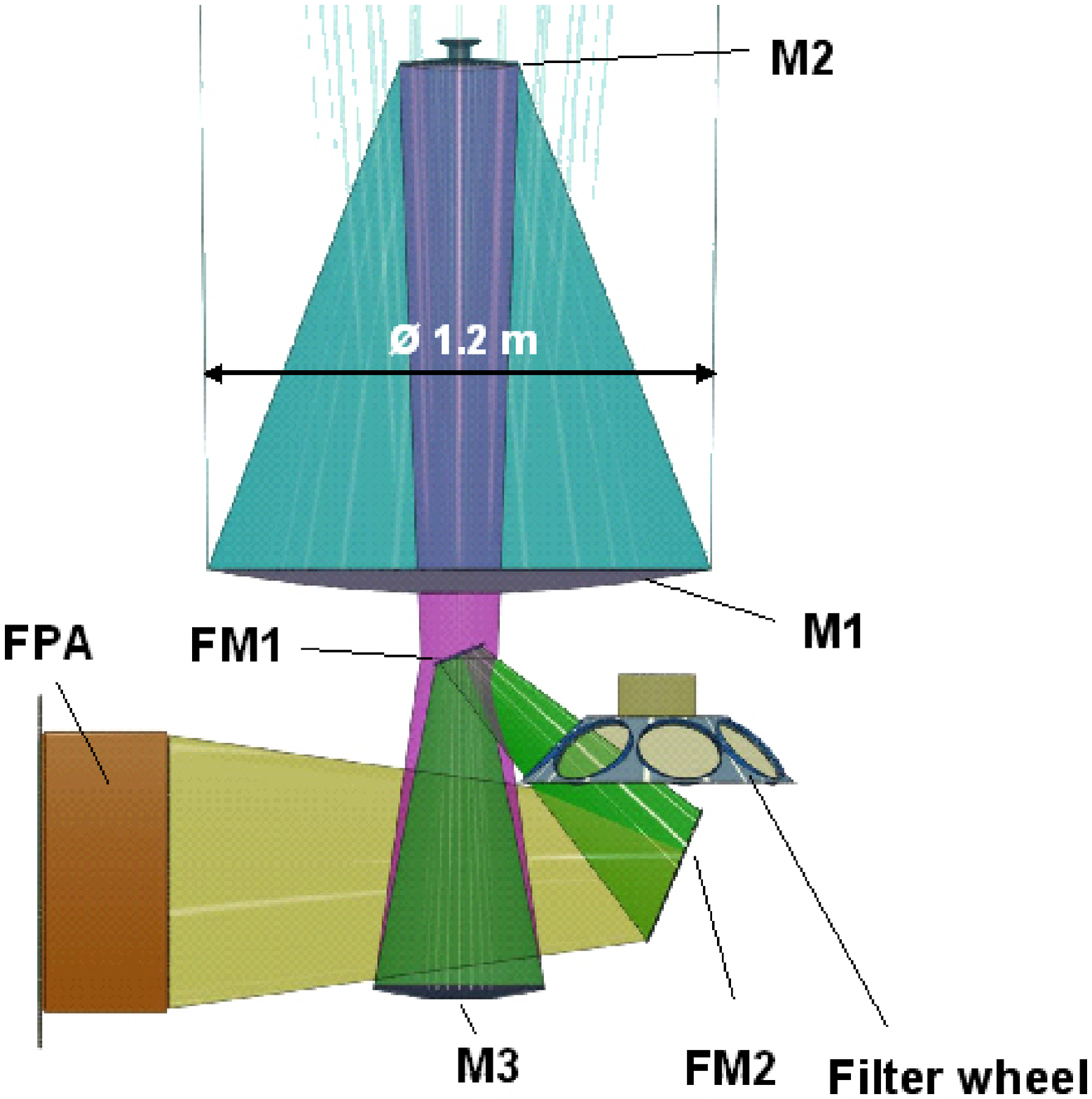}
\end{tabular}
\end{center}
\caption{\label{fig:optics}Baseline optical concept for the DUNE telescope. Left: NODI Optical design with an effective diameter of 1.2m (see ref. \citenum{gra06} for details).  Right: Schematic layout of the telescope, the filter wheel and the focal plane assembly.} 
\end{figure}

Figure~\ref{fig:optics} illustrates the optical design for a telescope with 2.46m total length
in the small folding flat configuration. A longer version would be harder to fit in the spacecraft but would have reduced sensitivity to misalignment of the secondary mirror. The optimization of the NODI design is made in such a way that along a full $0.12^{\circ}$ rotation of the instrument about the drift scan axis (ie 49.5 mm at the focal plane), all the optical images on all the CCD planes translate, at the same linear speed, along lines that are parallel to the columns of each CCD, with an RMS distortion on position less than 2$\mu$m. This is obtained at the cost of a slight curvature of the focal surface, a possibility that does not significantly alter the image quality or the complexity of the realization. A very tight alignment of each CCD along the local drift direction in the FOV is required (1$\mu$m at the edge), but the position in translation of the CCDs has relatively loose tolerances. A tight in-flight control of the drift direction and rotation speed is required (see \S\ref{aocs} below). 

\subsection{Focal plane assembly} 
\label{focal_plane}

\subsubsection{CCD detectors}
\label{ccd}

From the science requirements (\S\ref{requirements}), we see that the wavelength range that will be used spans the whole visible range. Specifically, we need detectors that have good quantum efficiency from 375nm to 950 nm. Two CCD detectors from e2v have been considered, namely the 4280 model and the 9172 model.
The 4280 series of CCDs is similar to the 4290 series that was used in the MegaCam camera built by CEA/DAPNIA\cite{ccds}. The 9172 series was developed specifically for the GAIA ESA mission. Given that the detection needs for GAIA are  close to those for DUNE (space application, observations in TDI mode, very low signals expected), the 9172 CCD is a natural choice for this mission.

The DUNE CCDs will need improved quantum efficiency in the red part of the visible spectrum, as compared to standard thinned backside illuminated devices: the 4280 CCDs can be readily ordered in their `deep depletion' version, while `red' 9172 CCDs have already been developed for GAIA. The typical quantum efficiencies for these devices 
is available from e2v\footnote{{\tt http://e2v.com/datasheets/publications/brochures/ccd\_selection\_guide.pdf}}.

Compared to the 4280 series, the 9172 series already has all the improvements needed for TDI operation at very low fluxes. The only modification to be done is on the pixel size, from $10 \times 30$  $\mu$m$^{2}$ in the case of the GAIA detectors, to $13.5 \times 13.5$ $\mu$m$^{2}$ in the case of DUNE. The design of new masks is therefore necessary. 
The 4280 series need more modifications, with the design of new masks to cope with the needs of DUNE: an additional 4th phase for the TDI mode, a charge injection mechanism to deal with the traps, and a serial register along the long side of the detector instead of the current short side.

Given the requirements on both the size and the sampling of the point spread function, the angular size of the pixel is chosen to be 0.115$''$. A 4280 CCD has $2048 \times 4096$ pixels and 60 of them will cover 0.51 deg$^{2}$. A GAIA CCD with a new pixel of $13.5 \times 13.5$ $\mu$m$^{2}$ will contain approximately $4355 \times 3330$ pixels, and 32 of them will cover 0.47 deg$^{2}$. The field of view can be covered either with 8 strips of 4 GAIA CCDs, or 10 strips of 6 4280 CCDs. In both cases, 4 additional devices are needed for the Attitude and Orbit Control System (AOCS).

\subsubsection{Signal levels}
\label{signal}
Both the weak lensing and the supernovae observing programs will lead to very low signals at the end of each CCD exposure. For the faintest objects in the surveys, the contributions of the sky background and the source will add up to roughly 60 e$^{-}$/pixel in the weak lensing case, and 6 e$^{-}$/pixel in the supernovae case at the level of the CCD output node. This is clearly not the optimal operating point for a CCD as far as output non linearity and charge transfer efficiency are concerned (see for example ref.~\citenum{acs_1} for performances of the HST ACS CCDs). 

For a constant integration time of 1500 seconds in TDI mode, the different sizes of the detectors translate into a higher number of 4280 CCDs along the scan direction than with the 9172 CCDs (6 instead of 4), and consequently a shorter exposure time per CCD (375 s for the GAIA CCD, 250 s for the 4280 CCD), and a low signal level reduced by 1/3 with the 4280 CCD. Since the DUNE and GAIA detection needs are similar, the GAIA type of CCD is thus again preferred to the 4280 type. A detailed assessment of the performances of the DUNE CCDs in this regime will be performed in the near future.

\subsubsection{Impact of Radiation}
\label{radiation}
For the Geosynchronous Earth Orbit (GEO) which is the baseline for the mission (see \S\ref{mission} below), the most damaging radiation impacting the CCD detectors are cosmic particles (mainly protons) and particles trapped in the van Allen belts (mainly electrons).
The first effect of this radiation is that the images are polluted by  the ``glitches'' induced by ionising energy deposited in the detector pixels. To evaluate the impact of this effect, we first consider a first order calculation that neglects secondary particles created by interactions between primary particles and the detector environment. Considering the pixel size, detector geometry, the solar cycle and models for cosmic protons, we find that, at the end of each 375 seconds exposure, roughly 0.6\% of pixels will be polluted by cosmic protons, which is negligeable. The same is not true for trapped particles for a spacecraft in a geosynchronous orbit. The trapped electrons are much less energetic but their flux is significantly higher than the flux of cosmic particles. While the low energy
electrons can be avoided with shielding, the most energetic electrons ($>$ 5 MeV) 
of the van Allen belts can still pass through a relatively thick shield (typically equivalent to 1 cm of aluminum). A detailed analysis, including the most recent radiation models  and an assessment of the performances of the `deglitching' algorithms, is crucial for assessing the impact of this effect and the required level of shielding.

The second impact of radiation is the ageing of the detectors. For CCD devices,  ageing  typically results in an increase in both the dark current and the number of hot pixels, and a loss in charge transfer efficiency. The latter effect can be problematic, since it affects the shape of the images of objects, which is the basis of the weak lensing measurements. 
Ageing is sensitive to many parameters and is strongly dependent on the technology used to manufacture the detectors. The effect has been clearly seen in the CCDs of the ACS camera (ref.~\citenum{acs_2}), while the CCDs developed for the GAIA mission seem to behave fairly well in this respect, laboratory tests are required to assess the extent and impact of this effect for DUNE. 

\subsubsection{Architecture}
\label{layout}

\begin{figure}
\centering
\includegraphics[width=60mm,height=65mm]{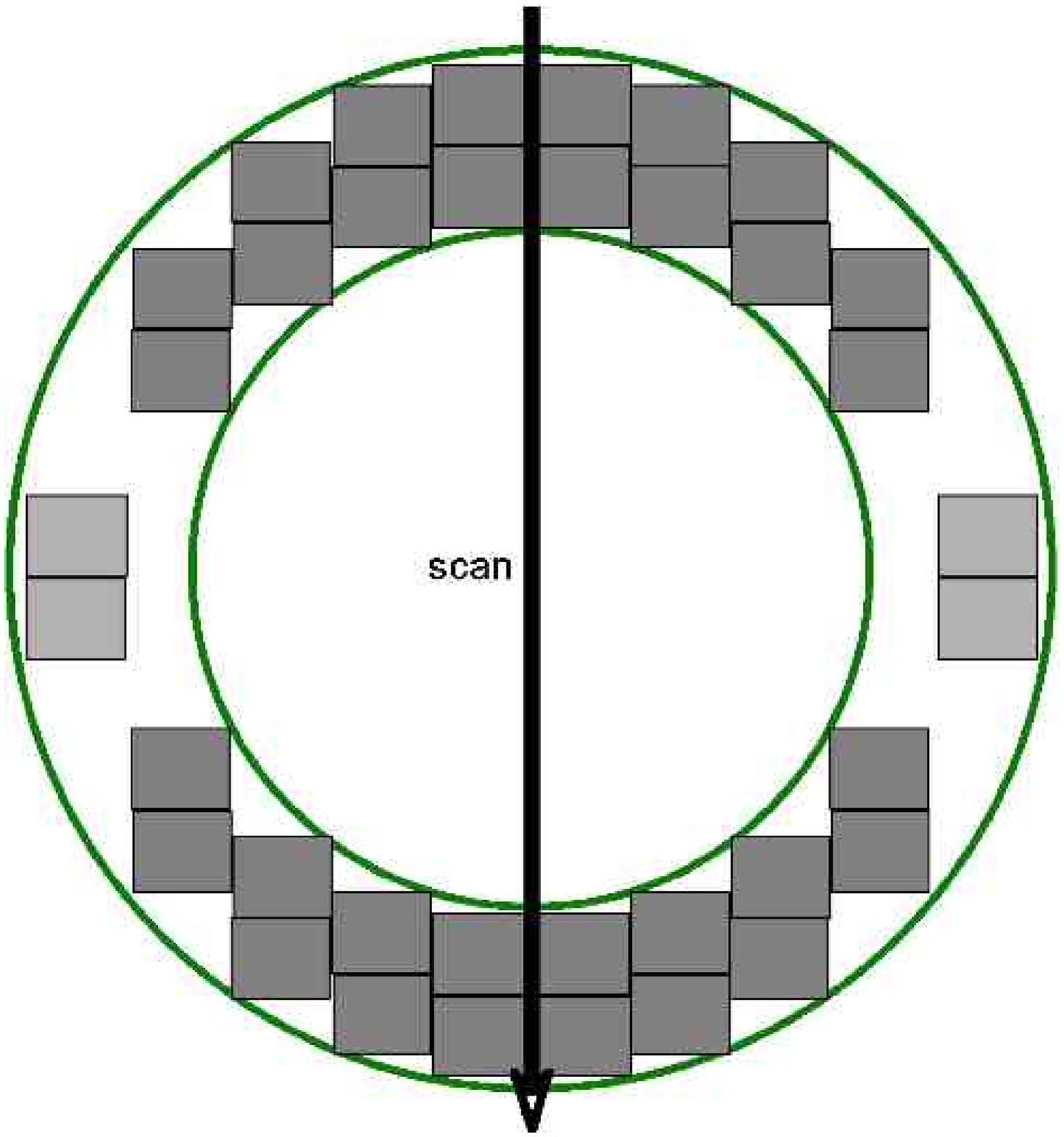}
\includegraphics[width=80mm,height=70mm]{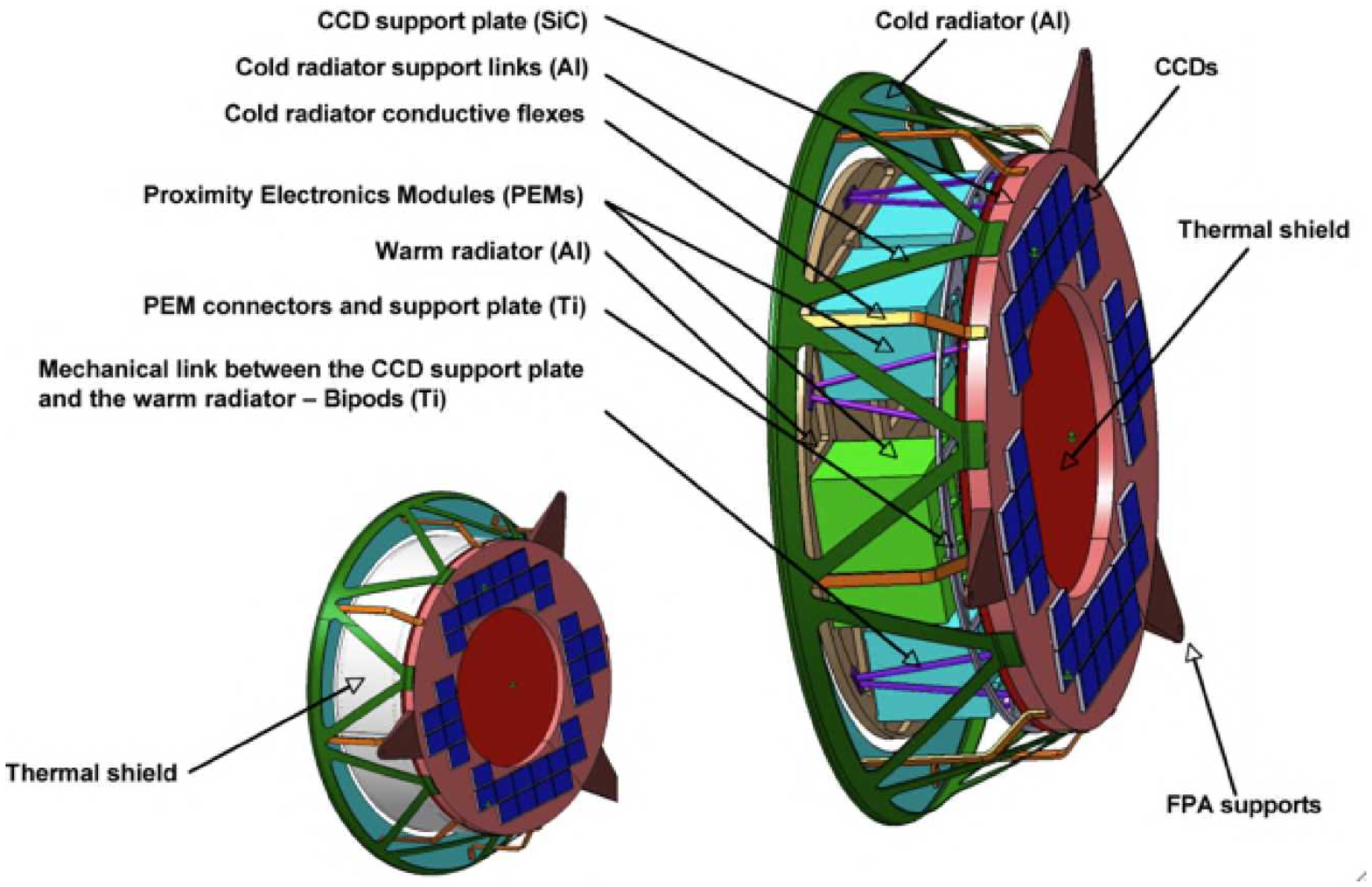}
\caption{Concept for the DUNE focal plane. Left: Focal plane layout with 
respect to the drift scan direction. The darker and lighter
blocks represent science and AOCS CCDs, respectively. Right:
Focal Plane architecture showing the CCD support plate, 
the electronics, the radiators and the mechanical and thermal support structure.
\label{fig:focal_plane}}
\end{figure}

As described in \S\ref{optics}, the optical design gives an annular field of view, with inner and outer radius of 208 and 320 mm, respectively. There is no specific requirement in the positioning of the CCDs along the scan direction, since the objects will cross each CCD
in a column. However, in order to have a complete coverage of the field of view across the scan direction, CCDs have to be slightly shifted from one row of detectors to the next, so that the gaps between two adjacent CCDs in a row are covered in the other rows. Figure~\ref{fig:focal_plane}a shows a possible layout of the 32 science and 4 AOCS GAIA CCDs.

Given the annular shape of the field of view, the basic mechanical concept is to have an annular support plate for the CCDs, and to place the front end electronics inside the annulus. The working temperature of CCDs is around 170K while the temperature
of the electronics is around 290K. Efficient thermal decoupling 
between the CCD support plate and the warm electronics is therefore needed. Two separate radiators will be used, one for the CCDs and one for the warm electronics, as shown in figure~\ref{fig:focal_plane}b. 

The front end electronics (FEE) needed to drive and read the CCDs will be distributed in a small number of units (typically 5 or 6), each unit being in charge of one or two strips of CCDs along the scan direction, rather than a group of neighboring devices. One of the main driver for the mission is the total sky area measured at the end of the mission, and the loss of only one CCD per strip means the loss of the full strip (as far as sensitivity is concerned); therefore losing one FEE controlling e.g. two CCDs on the same strip is less important than losing the control of two CCDs on two different strips, since the loss of sky coverage will be halved. The total power needed to drive all the CCDs is estimated to be around 150 W.

\section{Mission and Spacecraft}
\label{spacecraft}

\subsection{Mission Design}
\label{mission}
The design of the DUNE spacecraft is faced with stringent requirements on pointing stability during the observation time and on thermal stability of the focal plane. Both stability and thermal requirements impose restrictions on the selected operational orbit for the DUNE mission. The Geosynchronous Earth Orbit (GEO) is the baseline for the DUNE mission, providing a low perturbation environment and stable thermal conditions, as opposed to Earth orbits with lower altitude. As discussed in \S\ref{radiation}, specific tests on CCD detectors are required to evaluate the viability of the GEO orbit with respect to radiation.
A High Elliptical Orbit (HEO) of 253 000 km apogee and 44 000 km perigee (7 days period) is also a feasible option, provided small modifications on the telemetry system are made.

The DUNE spacecraft may be launched from Kourou using a direct injection by Soyuz ST-Fregat on the final GEO orbit. The spacecraft launch mass is 1140 kg, for an estimated launcher performance of 1200 kg (without adapter). If necessary, a dedicated propulsion module (recurrent from the Mars Express propulsion system) can be accommodated inside the spacecraft to bring performance margins. In that case, the injection will be done on a GTO+ transfer orbit (6500 km perigee altitude) by Soyuz ST-Fregat and a circularisation boost will be performed by the spacecraft itself at the apogee of the GTO+ orbit.

The spacecraft accommodates a Sun Shield  to protect the telescope and focal plane assembly from direct sun illumination (see figure~\ref{fig:spacecraft}a). The Sun Shield is a simple plane providing protection from $30^\circ$ sun aspect angle. Its normal direction is directed towards the Sun and the overall spacecraft is rotated around this axis to allow for the science scanning strategy. The angle between the Sun direction and the rotation direction (Sun Shield normal vector) can be tuned up to $30^\circ$, together with the lengths of the strips to be scanned each day. This flexibility allows optimal coverage of the sky, i.e. contiguous strips of data, one zone accessible every day. The rotation is made in scan-and-sweep-back approach where the spacecraft is brought back to next scan position after each strip. Each day, one strip is scanned and the resulting data are transmitted to ground during the attitude manoeuvre to get to next position

\subsection{Spacecraft Design}
\begin{figure}
\centering
\centerline{
\includegraphics[height=70mm]{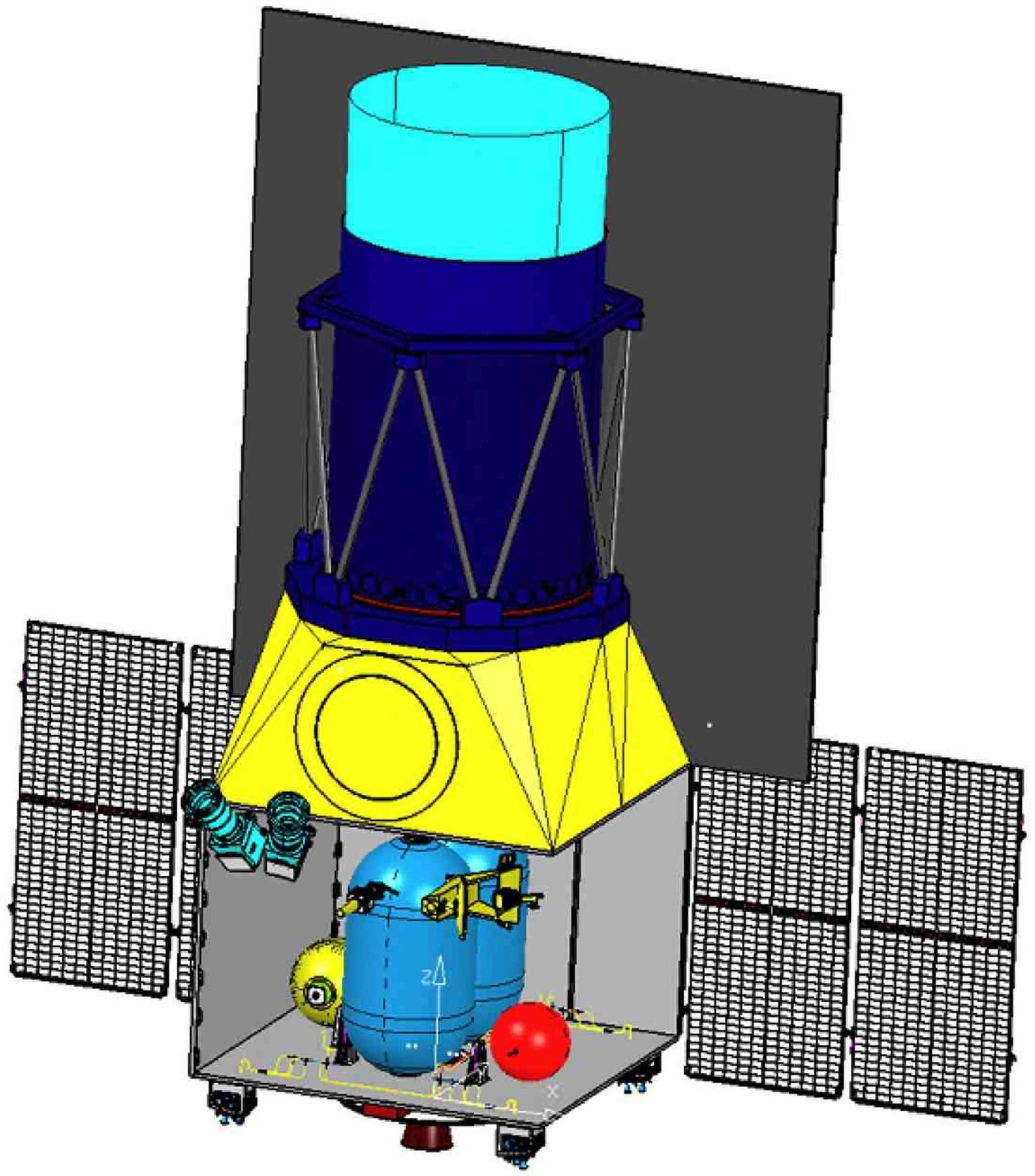}
\includegraphics[height=70mm]{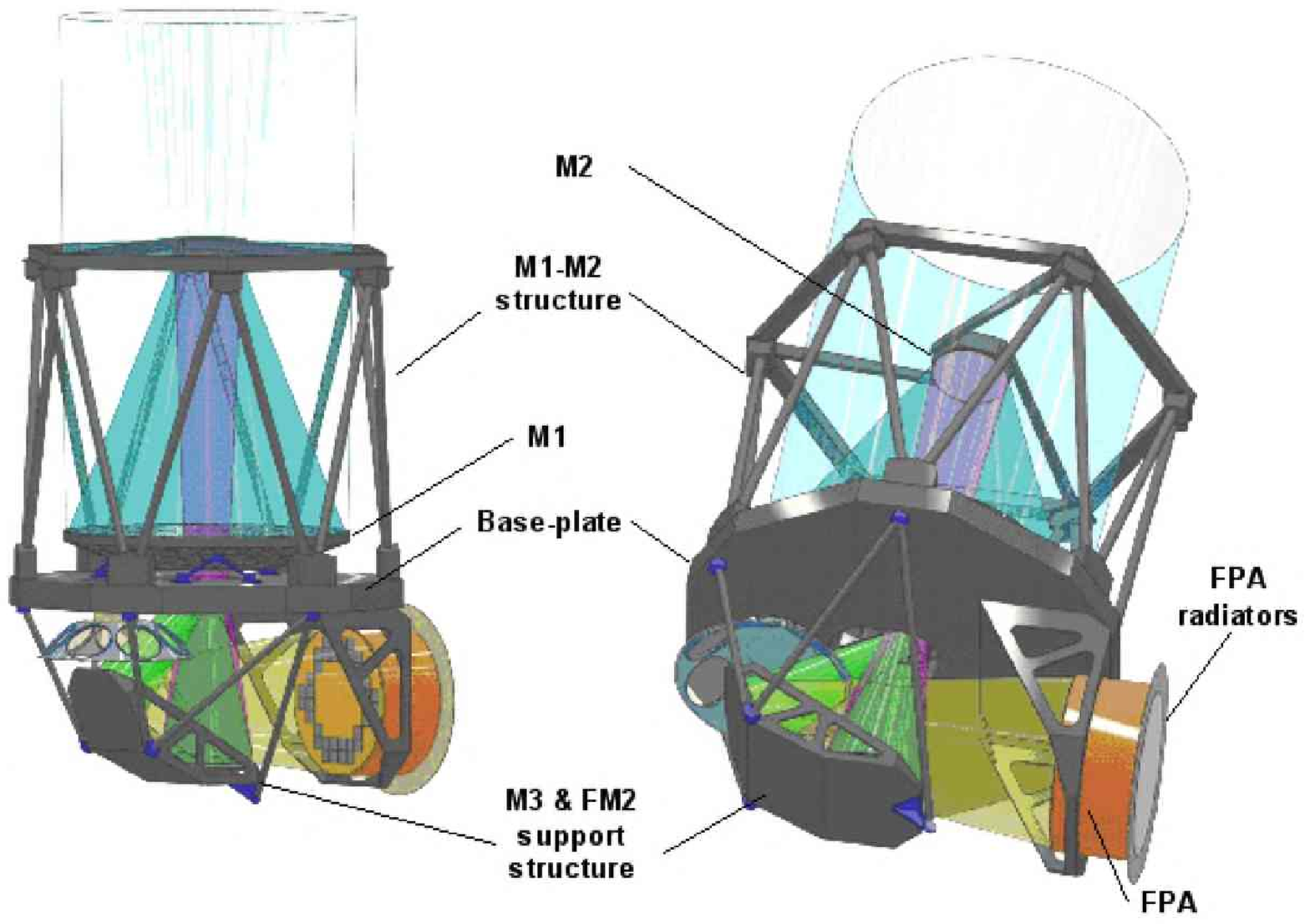}}
\caption{\small Spacecraft configuration with the Mars Express propulsion module (left). 
Payload Module with the telescope, focal plane assembly, filter wheel and SiC structure (right).
 \label{fig:spacecraft}}
\end{figure}

The spacecraft concept is shown in figure \ref{fig:spacecraft}a and consists of a Payload Module (PLM) that includes the instrument (telescope and focal plane assembly) and a Service Module (SVM).
The SVM design is based on the Mars Express structure and Solar Arrays (with modified cells). The Sun Shield is controlled by the service module. The telemetry chain is based on the existing Pleiades equipments (X-band telemetry, 155 Mbits/s modulator, one 800 Gbits Solid State Mass Memory) and a fixed 30 cm high gain antenna. Telemetry and telecommand (TM/TC) is provided through standard S-band chain with two low gain antennae.  A cold Gas micro-propulsion system is reused from the GAIA ESA program. A chemical propulsion system is used for coarse attitude control and manoeuvers. Two redundant Star Trackers are used with gyroscopes and sun sensors for the
AOCS. A recurrent Mars Express propulsion module can also be accommodated in the option of an injection by Soyuz-ST on a transfer orbit rather than the direct GEO injection scheme. The option of HEO operational orbit would change the high-gain antenna diameter (up to 70 cm used with 15 m ground station).

The payload module makes use of Silicon Carbide (SiC) structure (see figure \ref{fig:spacecraft}b).  The high mechanical and thermal properties of the SiC material make it an excellent candidate. It is also widely considered as excellent for the development of low mass large mirror substrates. 
It provides higher stiffness and less mass compared to classical CFRP/aluminium honeycomb structure and zerodur mirrors. Silicon Carbide will thus be used both for structural and mirrors parts to simplify the architecture and to ensure high stability and low mass. The processes of manufacturing and assembly have already been validated on many EADS Astrium instruments programs. The
SiC structure also supports the Focal Plane Assembly (described in \S\ref{focal_plane}) and the Filter Wheel for the Supernovae survey (see \S\ref{optics}).

\subsection{Spacecraft Pointing and Stability}
\label{aocs}

 The galaxy shape measurement in the weak lensing survey are effected by the overall `system' PSF, which includes contributions from the instrument (optics, detectors), from perturbations of the Attitude and Orbit Control System (AOCS) and from thermo-elastic deformations of the instrument. The spacecraft stability during the CCD integration time is therefore of prime importance for the spacecraft design. Preliminary evaluations showed that a direct allocation of the PSF stability requirement ($\delta e < 0.1$ percent, see \S\ref{requirements}) to the spacecraft AOCS  was prohibitive. Further studies have however shown that the PSF requirements could be met by sharing them between the AOCS and a posteriori ground based calibration. 

\begin{figure}
\centering
\includegraphics[width=120mm]{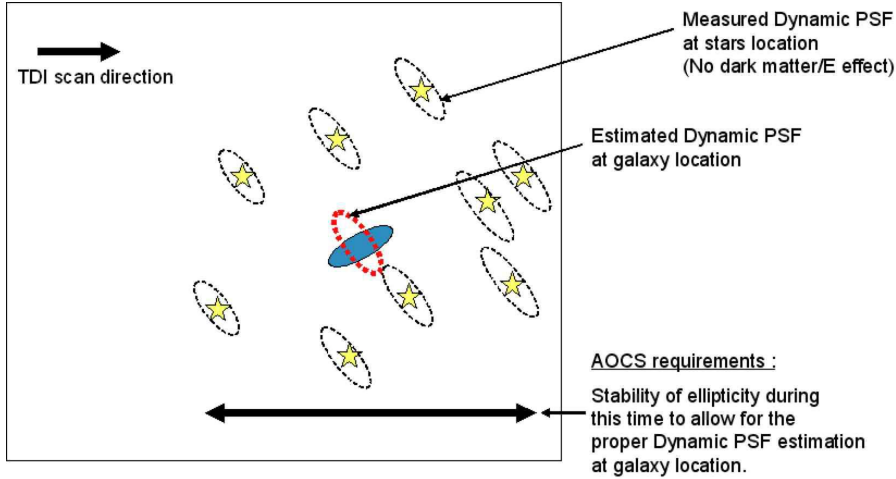}
\caption{\small Principle of shear measurement using calibration stars around a galaxy
to estimate and deconvolve the PSF \label{fig:calibration}}
\end{figure}

To reduce the constraints on the AOCS, bright stars in the neighborhood
of a galaxy will thus be used to estimate the shape of the system PSF at the galaxy location (see illustration in figure~\ref{fig:calibration}).  Each star gives a noisy estimation of the underlying PSF shape due to its finite SNR and to
pixelisation. This calibration process will therefore succeed only if the PSF shape is sufficiently stable over a region around the galaxy and if the number of stars is sufficient over this region. 
 
A preliminary study based on simulations with the system PSF for the DUNE baseline configuration has shown that between 40 and 120 stars are necessary to allow for adequate PSF estimation. 
Given a sampling rate and sample noise (from instrument properties such as TDI rate), the satellite design shall be such that the frequency and amplitude of the PSF shape parameter variation over a spatial dimension of the image (i.e. 40 to 120 stars) can be properly sampled.

The estimation of line-of-sight movements on the PSF shape variation over the image has been performed as a function of perturbation frequency and corresponding AOCS requirements has been derived. The driving requirement using this calibration method is the requirement on the PSF size itself (0.23 arcseconds) that calls for AOCS stability of 0.2 $\mu$rad over 375s, the integration time over a CCD length. This stability can be achieved using dedicated sensors on the focal plane to measure the spacecraft rates 
(see \S\ref{layout}), cold gas thrusters for accurate actuation as those used in GAIA, and hybridization of the Star Tracker with the payload to reduce the noise injected by the star tracker in the loop.

\section{Conclusions}
\label{conclusion}
The purpose of the DUNE mission is the high-precision study of dark matter, dark energy and the initial conditions of cosmic structure formation. The novel concept of DUNE relies primarily on the weak lensing technique which provides a direct probe of Dark Matter and the growth of cosmic structure. In addition, DUNE also includes a complementary Supernovae survey which will provide a measure of the expansion history of the universe. To achieve the high precision necessary for the study of dark energy, stringent control of systematics are built into the mission requirements. The mission concept elaborated during the CNES phase 0 study of DUNE remains to be optimised but was shown to meet the science requirements. In particular, a combination of AOCS performance with dedicated star tracking CCDs and of a posteriori PSF calibration was shown to provide the required control of the shear measurement systematics. 

In the current baseline, DUNE will provide a mono-chromatic wide field survey covering 20,000 deg$^{2}$ optimised for weak lensing, and a deep 120 deg$^2$ survey in 6 bands for the Supernovae. These surveys together with the tight control on systematics
will allow DUNE to measure the equation of state parameter of the Dark energy and its evolution with a precision of a few percent and a few tens of percent, respectively. The size and image quality of these surveys will also provide a wealth of secondary science returns such as the study galaxy formation, cross-correlation with CMB foregrounds, the construction of a mass selected cluster catalogue, the history of star formation using Type II supernovae, etc. The multi-color deep survey will also allow us to produce a full high-resolution 3D mass and light reconstruction. 

The philosophy of the DUNE mission is that of a fast mission with limited risks and costs. To reduce the costs, the DUNE instrument relies, where possible, on existing components and makes use of ground based survey complements. This ground/space synergy ensures that only observations for which the ground is insufficient are performed in space. 

In addition to the ongoing CNES study, DUNE is also being proposed to ESA's Cosmic Vision programme. In this context, further studies of the mission concept will take place, both to further the feasibility of the mission and to reevaluate its scope. In particular, 
further studies of the ground/space synergy will determine whether the wide survey
will provide more bands in space, thus improving the accuracy of photometric redshifts. CCD radiation experiments will be performed to assess the performance 
of the detectors in the baseline GEO orbit. In the context of an ESA mission, the extension of the wavelength coverage of DUNE into the NIR will also be studied, along with a moderate increase of the telescope diameter.

\acknowledgements
We are  grateful to Pierre-Olivier Lagage, Laurent Vigroux, Francis Bernardeau and Olivier Lef\`{e}vre for their suggestions and support. We thank Nabila Aghanim, Jean-Loup Puget and Fabienne Casoli at the Institut d'Astrophysique Spatial in Orsay, who recently joined the
collaboration for useful suggestions and discussions. We are also grateful to Jason Rhodes and Anne Ealet for useful discussions. This DUNE project is supported by CNES, CEA, CNRS, the Programme National de Cosmologie and the Programme Astroparticules.

%%%%%%%%%%%%%%%%%%%%%%%%%%%%%%%%%%%%%%%%%%%%%%%%%%%%%%%%%%%%%
%%%%% References %%%%%

%\bibliography{dune}   %>>>> bibliography data in dune.bib
%%\bibliography{report}
%\bibliographystyle{spiebib}   %>>>> makes bibtex use spiebib.bst

\end{document}